\documentclass[twocolumn,secnumarabic,amssymb, nobibnotes, aps, prd]{revtex4-2}

 \clearpage

\def\psl{\hbox{\hbox{${p}$}}\kern-1.9mm{\hbox{${/}$}}}
\def\dsl{\hbox{\hbox{${\partial}$}}\kern-1.7mm{\hbox{${/}$}}}
\def\Dsl{\hbox{\hbox{${D}$}}\kern-2.1mm{\hbox{${/}$}}}

\newcommand{\bp}{M_P}

\usepackage{epstopdf}

\def\be{\begin{equation}}
\def\ee{\end{equation}}
\def\bea{\begin{eqnarray}}
\def\eea{\end{eqnarray}}

\def\ba{\begin{array} }
\def\ea{\end{array}}
\def\bac{\begin{array} {c}}
\def\bacc{\begin{array} {cc}}
\def\baccc{\begin{array} {ccc}}

\usepackage{amssymb}
\usepackage{amssymb,exscale}
\usepackage{color}
\usepackage{xcolor}
\usepackage{mathrsfs}
\usepackage{epsfig}
\usepackage{mathrsfs}
\usepackage{amsfonts}
\usepackage{slashed}
\usepackage{blindtext}
 \usepackage{fancyhdr}
 \usepackage{mathtools}
 
\usepackage{comment}
\usepackage{graphicx}
\usepackage{amsmath}
\usepackage{mathtools}
\usepackage{slashed}
\usepackage{xcolor}
\usepackage[force]{feynmp-auto}
\usepackage{grffile}
\usepackage{float}
\usepackage{bbm}
\usepackage{hyperref}
\usepackage{ragged2e}
\usepackage{blindtext}

\begin{document}

\title{\LARGE Independent connection in ACTion during inflation}%

\author{{\large Alberto Salvio} \\
\vspace{0.2cm}
{\it Physics Department, University of Rome and INFN Tor Vergata, Italy}\\
\vspace{0.6cm}
\begin{abstract}
\noindent The Atacama Cosmology Telescope (ACT) has recently released new measurements and constraints on inflationary observables. In this paper it is shown that a component of a dynamical affine connection, which is independent of the metric, can easily drive inflation in agreement with these observations.  Such geometrical explanation of inflation is analysed in detail here in the minimal model, including the predictions for the scalar spectral index $n_s$ and its running $\alpha_s$, the amplitude of the scalar perturbations and the tensor-to-scalar ratio $r$. Furthermore, this minimal model is shown to provide an inflationary attractor: arbitrary initial values of the kinetic energy density are  dynamically attracted down to negligible values compared to the potential energy density in homogeneous and isotropic metrics. Also, the role of the Higgs boson during and after inflation is briefly discussed.
\end{abstract}
}%



\maketitle


\section{Introduction}

\noindent  In Refs.~\cite{ACT:2025fju,ACT:2025tim} the ACT collaboration has provided new determinations of several inflationary observables. One of the most evident differences with respect to the values previously furnished by the Planck and BICEP/Keck (BK18) collaborations~\cite{Ade:2015lrj,BICEP:2021xfz}  is a higher favored value for $n_s$: when the new ACT observations are combined with Planck data and baryon acoustic oscillation data from the Dark Energy Spectroscopic Instrument (DESI)~\cite{DESI:2024uvr} (this combination is denoted P-ACT-LB) Ref.~\cite{ACT:2025fju} finds $n_s=0.9743 \pm 0.0034$. 

This new result has attracted much attention~\cite{Kallosh:2025rni,Aoki:2025wld,Berera:2025vsu,Dioguardi:2025vci,Brahma:2025dio,Gialamas:2025kef} as several inflationary models that were in perfect agreement with the previous Planck and BICEP/Keck results are in tension with the new determination. 

The main purpose of this paper is to offer a well-motivated scenario that can explain the new results of Refs.~\cite{ACT:2025fju,ACT:2025tim}.
While Einstein's general relativity (GR) explains gravity in geometrical terms (the distances are measured through the metric and the gravitational force is determined by the affine connection), inflation is usually accounted for by adding an ad hoc scalar field, which is unrelated to geometry. However, from the purely geometrical point of view the metric and the connection, unlike in GR, can be completely independent objects and, moreover, can contain extra degrees of freedom besides the spin-2 graviton. This generalized scenario is known as metric-affine gravity (see~\cite{Baldazzi:2021kaf,Pradisi:2022nmh} for detailed discussions and further references). The extra degrees of freedom carried by the affine connection can include the inflaton~\cite{Salvio:2022suk,DiMarco:2023ncs,Karananas:2025xcv} and this provides a geometrical origin also for this scalar field and the subsequent period of reheating too~\cite{Salvio:2022suk,DiMarco:2023ncs}, which sets the initial conditions for the radiation dominated phase of the universe.

Addressing the compatibility of such scenario with the new ACT results  is not the only purpose of this work. Other new goals  include studying the implications of the couplings between the inflaton and the Standard Model particles for reheating and inflation, investigating the presence of an inflationary attractor and  studying the running of the spectral index in this scenario.

\section{The model}

\noindent An explanation of the new results of Refs.~\cite{ACT:2025fju,ACT:2025tim} can already been obtained in the simplest model where the inflaton is a dynamical scalar component of the affine connection (see~\cite{Salvio:2022suk,Pradisi:2022nmh} for an introduction to this model). The part of the action that is relevant for inflation in this case is 
\be S_I= \int d^4x\sqrt{-g}\left( \alpha{\cal R}+\beta {\cal R'} + c {\cal R'}^2\right), \label{SI}\ee
where~\cite{indices} ${\cal R}\equiv {\cal R}_{\mu\nu}^{~~~\mu\nu}$ and  ${\cal R'}\equiv \epsilon^{\mu\nu\rho\sigma}{\cal R}_{\mu\nu\rho\sigma}/\sqrt{-g}$ is the parity-odd Holst invariant~\cite{Hojman:1980kv,Nelson:1980ph,Holst:1995pc}. Here 
\be {\cal R}_{\mu\nu~~\sigma}^{~~~\rho} \equiv \partial_\mu{\cal A}_{\nu~\sigma}^{~\,\rho}+{\cal A}_{\mu~\lambda}^{~\,\rho}{\cal A}_{\nu~\sigma}^{~\,\lambda}- (\mu\leftrightarrow \nu)\ee
is the curvature built with the affine connection ${\cal A}_{\mu~\sigma}^{~\,\rho}$, also 
$\epsilon^{\mu\nu\rho\sigma}$ is the totally antisymmetric Levi-Civita symbol with $\epsilon^{0123}=1$ and $g$ is the determinant of the metric. The invariants $\mathcal{R}$ and $\mathcal{R}'$ generically appear in low-energy metric-affine effective field theories~\cite{Racioppi:2024zva}. In the GR case, where ${\cal A}_{\mu~\sigma}^{~\,\rho}$ equals the Levi-Civita connection, ${\cal R}$ coincides with the Ricci scalar, $R$, but ${\cal R'}$ vanishes. Thus in metric-affine gravity ${\cal R'}$ can be understood as a component of the connection.  The coefficients $\alpha$ and $\beta$ have dimensions of energy squared, while $c$ is a dimensionless coefficient. The quantity $\bp^2/(4\beta)$ is called the  Barbero-Immirzi parameter~\cite{Immirzi:1996di,Immirzi:1996dr}. These coefficients can be determined as follows. 

After introducing an auxiliary field $\omega$, which trades the curvature-squared term in~(\ref{SI}) with a linear-in-curvature term, and solving exactly the ${\cal A}_{\mu~\sigma}^{~\,\rho}$ field equations, one obtains the equivalent form
 \be S_I = \int d^4x\sqrt{-g}\left[\alpha R-\frac{(\partial \omega)^2}{2} -\mathcal{U}(\omega) \right],\label{SeqCan}\ee
 where
 \be \mathcal{U}(\omega)\equiv\frac{\left(\beta -\frac{\alpha}{2}  \sinh \left(X(\omega)\right)\right)^2}{4 c}\label{Uofo}\ee
 and
 \be X(\omega)\equiv \frac{\omega}{\sqrt{3\alpha}}+\tanh ^{-1}\left(\frac{4 \beta }{\sqrt{16 \beta ^2+4\alpha^2}}\right). \ee 
 Thus, one reproduces GR coupled to an ordinary scalar for $\alpha=\bp^2/2$, where $\bp$ is the reduced Planck mass. So, from now on $\alpha=\bp^2/2$. The coefficients $c$ and $\beta$ can be determined by requiring $\mathcal{U}$ to be a good inflationary potential. 
 
  The field $\omega$, the candidate inflaton, has the same symmetry properties of $\mathcal{R}'$~\cite{Hecht:1996np,BeltranJimenez:2019hrm,Pradisi:2022nmh}, so it is a spin-0 field with odd parity. For this reason it is named the pseudoscalaron. The squared mass of $\omega$ can be obtained by evaluating the second derivative of $\mathcal{U}$ at $\omega =0$, which is the only stationary point of $\mathcal{U}$. The result is 
 \be m^2_\omega = \frac{16 \beta ^2+\bp^4}{48 c \bp^2}. \ee
 Therefore, requiring the absence of tachyons, $c>0$. Also note that $\mathcal{U}(\omega)$ is symmetric in the exchange $\{\omega, \beta\} \to \{-\omega, -\beta\}$. Thus, an arbitrary value of $\beta$ and its opposite are physically equivalent. 
 
 For $\sqrt{|\beta|}\gtrsim \bp$, a plateau appears in $\mathcal{U}$ (see the blue line in Fig.~\ref{ps}), which is suitable for inflation.
 
 Despite its simplicity, this minimal model predicts inflationary observables that are quite robust under extensions of the inflationary action in~(\ref{SI}). First,  the same inflationary predictions emerge if one substitutes $c{\cal R'}^2$ with a general quadratic function of both ${\cal R}$ and ${\cal R'}$. This is because such more general function leads to the same potential~\cite{Pradisi:2022nmh} (see also~\cite{Gialamas:2022xtt}). Second, a variation of this idea where the action includes {\it only} a quadratic function of both ${\cal R}$ and ${\cal R'}$, and so it is  invariant under $g_{\mu\nu}\to \Omega^2 g_{\mu\nu}$, for a generically spacetime dependent $\Omega$ (Weyl invariance),   leads again to the same potential~\cite{Karananas:2025xcv}.

\section{Inflationary predictions}

The slow-roll approximation can be used when 
\be \epsilon \equiv\frac{\bp^2}{2} \left(\frac{1}{\mathcal{U}}\frac{d\mathcal{U}}{d\omega}\right)^2\ll 1, \quad \eta \equiv \frac{\bp^2}{\mathcal{U}} \frac{d^2\mathcal{U}}{d\omega^2}\ll 1. \label{epsilon-def}
\ee
If so, the number of e-folds $N_e$ as a function of $\omega$ is 
  \be N_e(\omega) = N(\omega)  - N(\omega_{\rm end}),    \ee
  where  
  \be N(\omega)  = \frac1{\bp^2}  \int^\omega d\omega'  \,  \mathcal{U}\left(\frac{d\mathcal{U}}{d\omega'}\right)^{-1}  \ee 
and $\omega_{\rm end}$ is the field value where inflation ends. The quantities $n_s$, $r$ and the curvature power spectrum $P_R$ (at horizon exit) are then given by
\be n_s = 1- 6\epsilon +2\eta, \quad r =16\epsilon, \quad P_R= \frac{\mathcal{U}/ \epsilon}{24\pi^2 \bp^4} \label{ns-r-PR}.\ee
 One finds analytic expressions not only for $\epsilon$, $\eta$, $n_s$, $r$ and $P_R$, but also for the e-fold function $N$ and $\omega_{\rm end}$, so $N_e$ can be computed analytically in terms of the generic field value $\omega$ (these analytic expressions can be found in~\cite{Salvio:2022suk}).

 Here we are also interested in predicting $\alpha_s$ (at horizon exit), which in the slow-roll approximation is
\be \alpha_s  = 16 \epsilon \eta - 24 \epsilon^2- 2 \xi_2, \label{eq:alphas} \ee
where the extra slow-roll parameter $\xi_2$ reads
\be \xi_2 \equiv  \frac{M_P^4}{\mathcal{U}^2}\frac{d\mathcal{U}}{d\omega} \frac{d^3\mathcal{U}}{d\omega^3}.\ee
Using the explicit expression of $\mathcal{U}$ one finds
\be \xi_2(\omega) = \frac{64 \bp^4 \cosh ^2(X(\omega)) \left(\bp^2 \sinh (X(\omega))-\beta \right)}{9 \left(\bp^2 \sinh (X(\omega))-4 \beta \right)^3}. \nonumber \\
\ee

Before comparing these inflationary predictions with observational bounds let us recall that, in using the slow-roll approximation, one is implicitly assuming the initial condition for the field momentum $\Pi\equiv\dot\omega$, where a dot is the cosmic-time derivative, to be small compared to the square root of the potential energy density, at least for the relevant values of $\omega$ on the inflationary plateau. In general one wants to explain why this is a reasonable assumption. In the present model, for homogeneous and isotropic metrics~\cite{HomIso}, initial values of the kinetic energy density $\Pi^2$ of order of, or even much larger than, $\mathcal{U}$ are dynamically attracted to small values, where the above-mentioned assumption is automatically verified. 

\begin{figure}[t]
\begin{center}
 \includegraphics[scale=0.39]{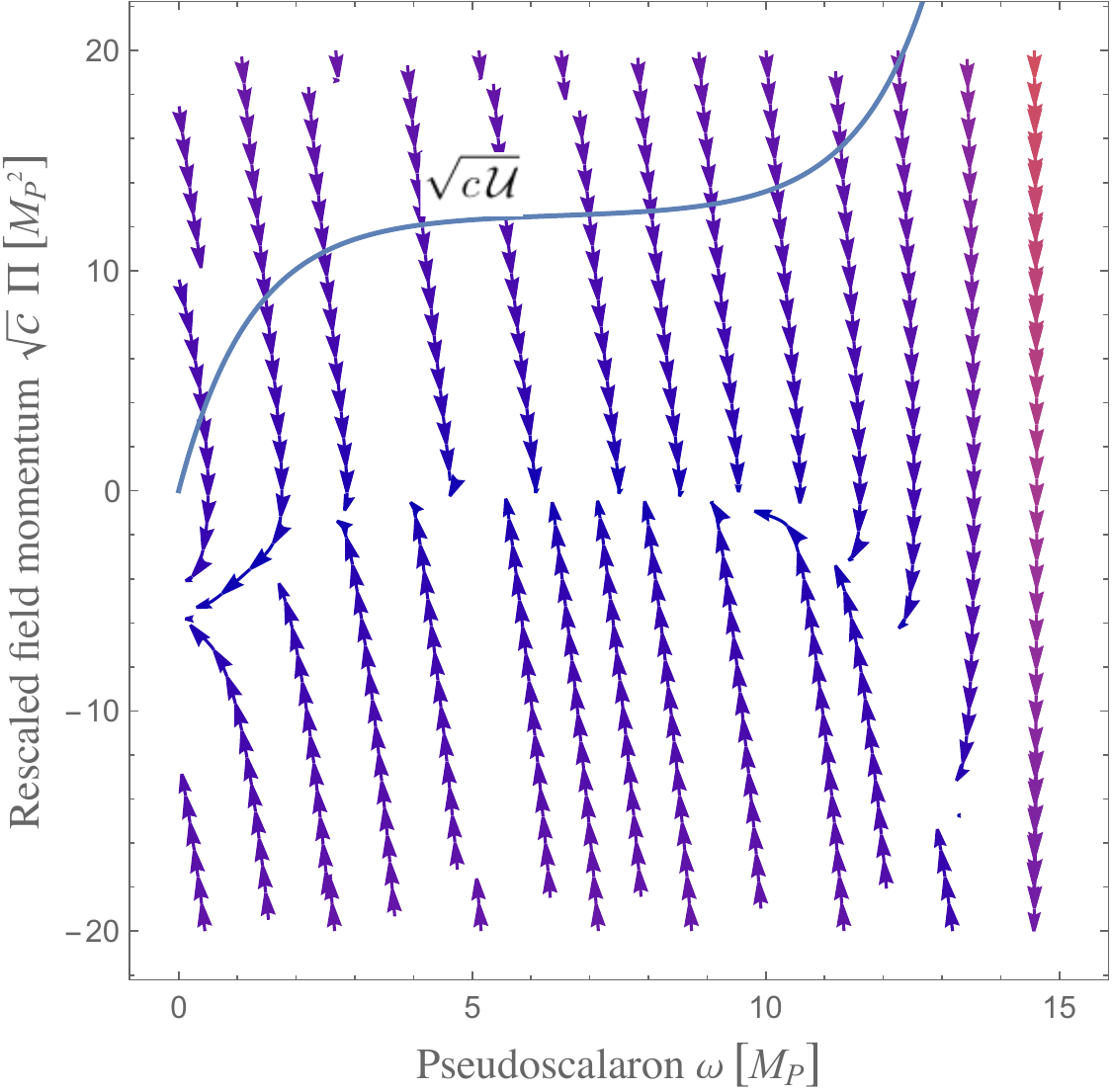} 
 \end{center} 
 \vspace{-.5cm}
   \caption{\em Phase space of the inflationary dynamical system in~(\ref{psS}) together with the square root of the potential energy density, $\sqrt{\mathcal{U}}$, for comparison. The arrows indicate the directions of motion. The field momentum $\Pi$ and $\sqrt{\mathcal{U}}$ have been multiplied by $\sqrt{c}$ to obtain a $c$-independent plot. Here $\beta=-25\bp^2$, but for other values of $\beta$ a qualitatively similar situation is obtained.
   }
\label{ps}
\end{figure}

To show this, first note that for inflationary actions of the form~(\ref{SeqCan}), with $\mathcal{U}$ not necessarily given by Eq.~(\ref{Uofo}), initial values of $\Pi^2$ much larger than $\mathcal{U}$ are always dynamically brought at least down to values comparable to $\mathcal{U}$~\cite{Linde:1985ub,Salvio:2015kka,Salvio:2017oyf}. There is no model-independent argument showing that $\Pi^2$ continues to decrease to values much smaller than the potential energy density, so the explicit form of $\mathcal{U}$ should be considered. The Einstein and inflaton field equations imply the following dynamical system
\be  \bac  
\left\{\bac \Pi'  =- 3\Pi-\frac1{H}\frac{d\mathcal{U}}{d\omega},  \\  \\ \hspace{-1.7cm}  \chi'= \frac{\Pi}{H}, \label{psS}
\\  \\ \hspace{-1.cm}H^2=\frac{\Pi^2+2\mathcal{U}}{6 \bp^2},
 \ea \right.  \ea \ee
where a prime is a derivative with respect to $N_e$  and  the spatial curvature has been neglected as the energy density is expected to be dominated by the inflaton during inflation. Fig.~\ref{ps} shows the phase space of this system with the directions of motion, illustrating how, even starting from values of $\Pi^2$ that are not small compared to $\mathcal{U}$, the dynamics brings the system to negligible values of $\Pi$ for $\omega$ on the inflationary plateau. Slow-roll then takes over as will be shown in Sec.~\ref{Facing observational bounds}.

\section{Facing observational bounds}\label{Facing observational bounds}

Note that, as always the case in slow-roll inflation, $\epsilon$, $\eta$ and $\xi_2$ (so also $n_s$, $\alpha_s$ and $r$) are independent of the overall constant in the potential, while $P_R$ is proportional to it. This overall constant is $1/c$ in the present case, see Eq.~(\ref{Uofo}). So, the observed value of $P_R$ (Ref.~\cite{Ade:2015lrj} gives $(2.10 \pm 0.03) \times 10^{-9}$)  can always be obtained by choosing $c$ appropriately.

\begin{figure}[t]
\begin{center}
\hspace{-0.37cm} 
 \includegraphics[scale=0.41]{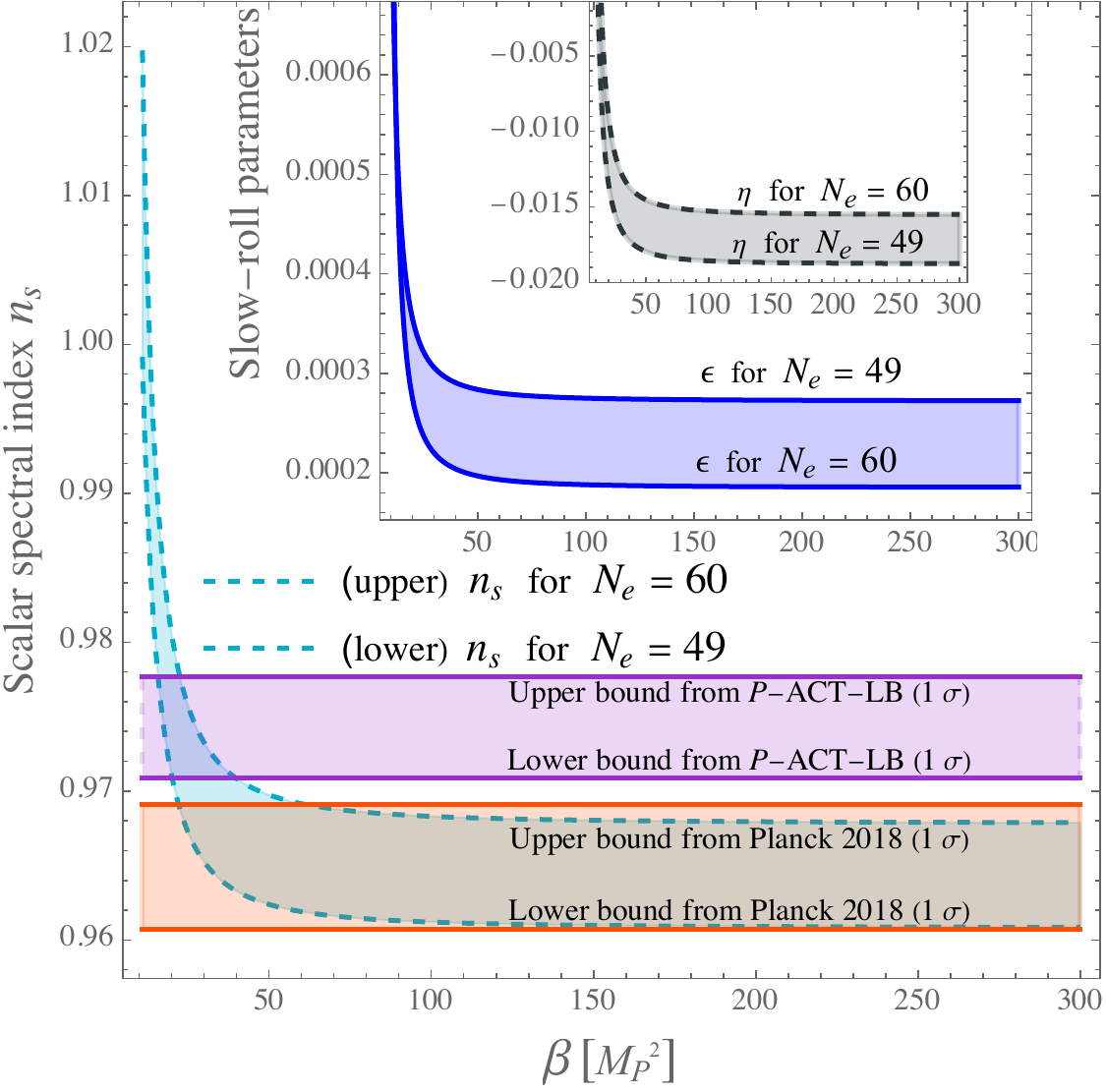} 
 \\ 
 \vspace{0.46cm}
 \includegraphics[scale=0.41]{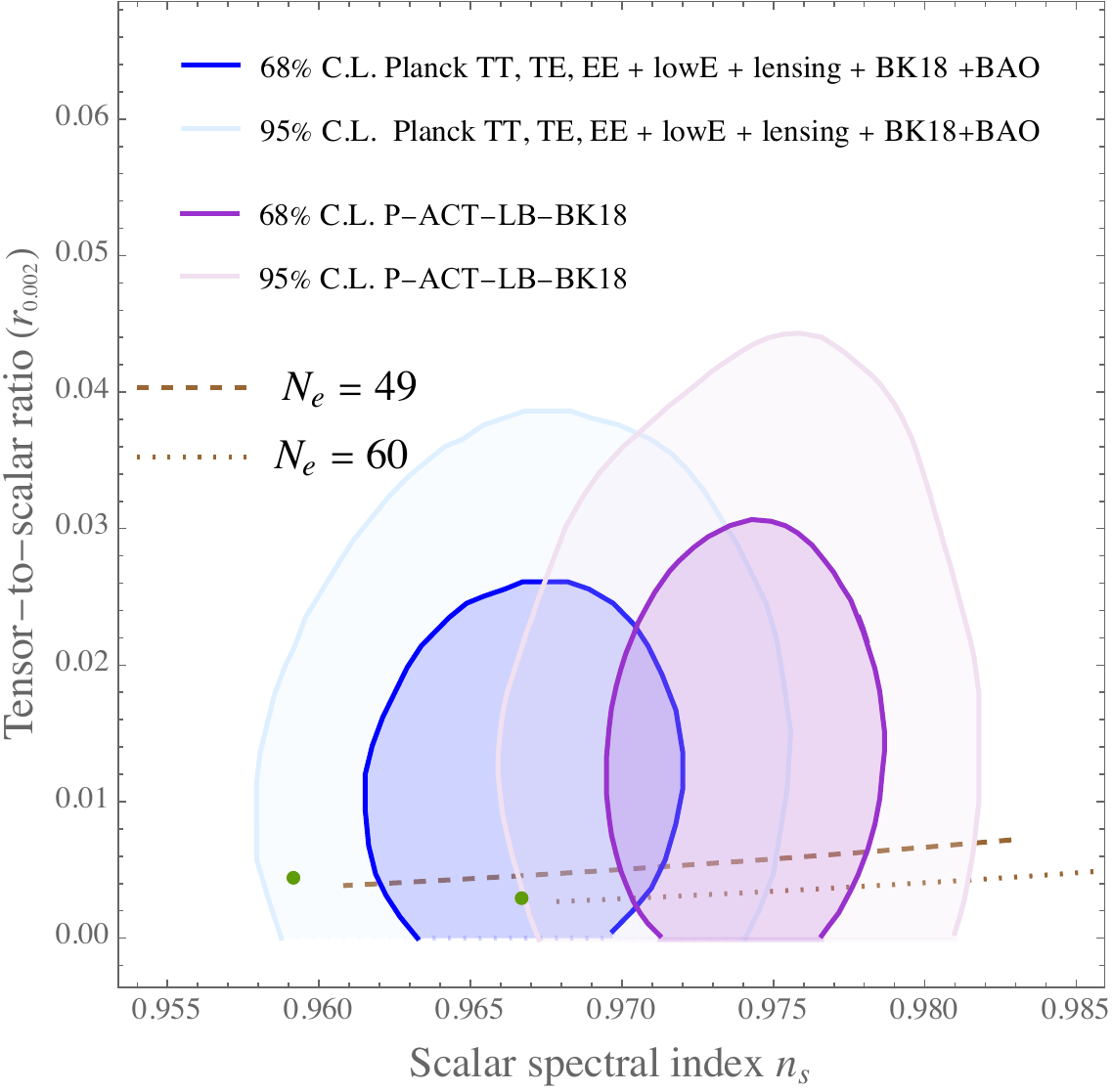}  \\
 \end{center} 
  \vspace{-0.5cm}
   \caption{\em {\bf Upper plot}: the scalar spectral index $n_s$ predicted by pseudoscalaron inflation as a function of $\beta$ together with the $1\sigma$ bounds from P-ACT-LB and Planck 2018 (the second paper in~\cite{Ade:2015lrj}). The inset shows the slow-roll parameters $\epsilon$ and $\eta$. {\bf Lower plot}:  the predictions of pseudoscalaron inflation in the $(n_s,r)$ plane together with the bounds from~\cite{Ade:2015lrj,BICEP:2021xfz} (in blue) and  the more recent ones from~\cite{ACT:2025fju,ACT:2025tim}. The green dots in the bottom plot are the predictions of Starobinsky inflation~\cite{Starobinsky:1980te}.}
\label{inflation}
\end{figure} 

In Fig.~\ref{inflation} it is shown  that slow-roll inflation not only occurs, but can also  remarkably accomodate the recent ACT bounds~\cite{ACT:2025fju,ACT:2025tim} for an appropriate number of e-folds $N_e$. In  Fig.~\ref{inflation}  the previous Planck and BK18 bounds of Refs.~\cite{Ade:2015lrj,BICEP:2021xfz} are also shown. In that figure $r_{0.002}$ is the value of $r$ at the reference momentum scale $0.002~{\rm Mpc}^{-1}$, used by Planck  and BK18. One can note that the ACT results~\cite{ACT:2025fju,ACT:2025tim} favor  values of $\sqrt{|\beta|}$ closer to $\bp$ compared to the Planck and BK18 bounds of Refs.~\cite{Ade:2015lrj,BICEP:2021xfz}. Values of $\sqrt{|\beta|}$ closer to the other dimensionful scale in the model, $\bp$, appear to be a more natural setting. For example, the minimal values of $\sqrt{|\beta|}$ to stay in the $1\sigma$ region of ~\cite{ACT:2025fju,ACT:2025tim} for $N_e =60$ and $N_e = 49$, respectively, are approximately $4\bp$ and $3\bp$, both of order $\bp$. The inset in the upper plot of Fig.~\ref{inflation} shows that for these values of $\beta$ the slow-roll approximation is still valid. 

\begin{figure}[t]
\begin{center}
 \includegraphics[scale=0.41]{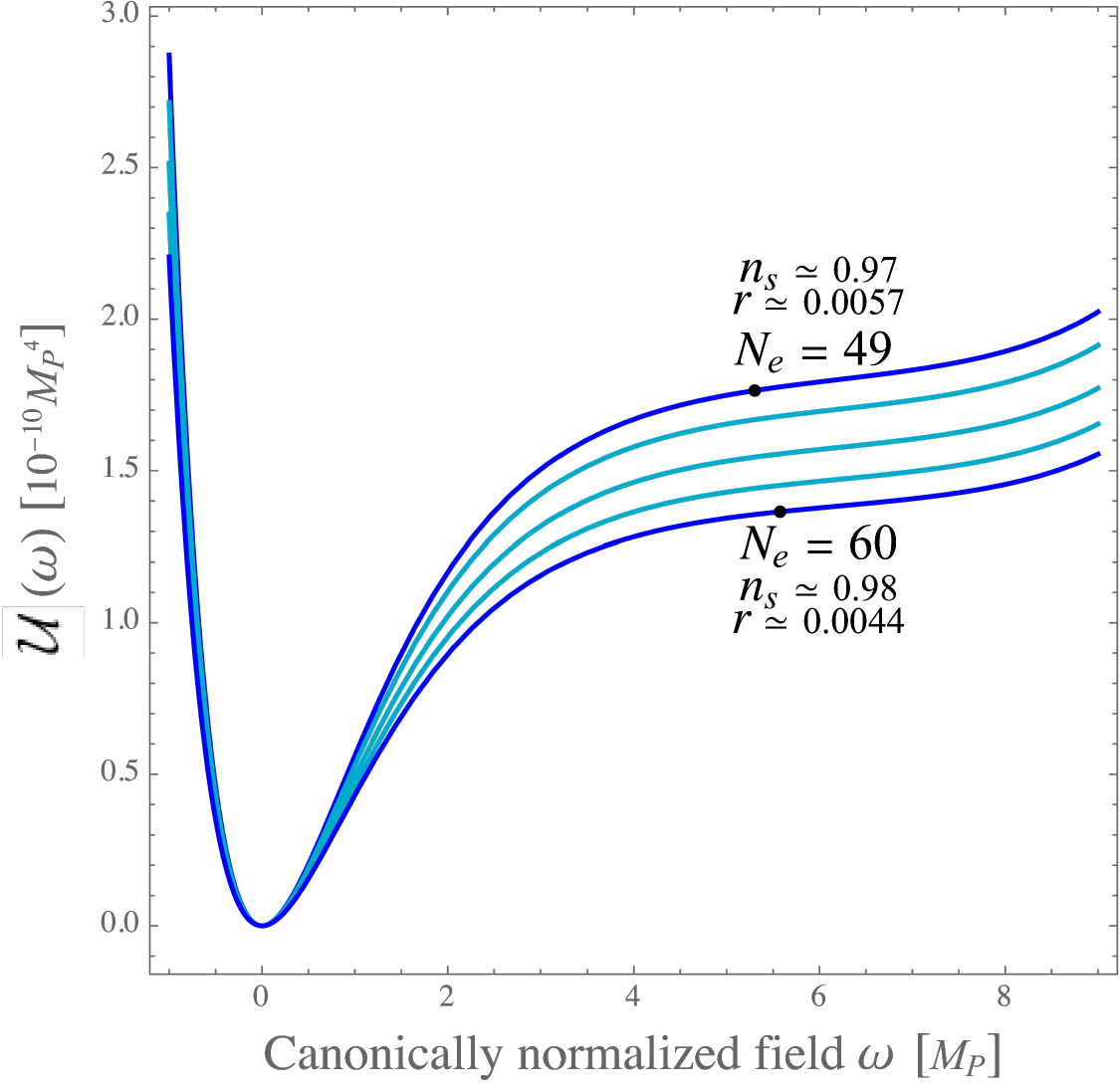} 
 \end{center} 
  \vspace{-0.5cm}
   \caption{\em The potential density as a function of the canonically normalized field. Here $\beta=-20\bp^2$.  For all curves $c$ is fixed to reproduce $P_R=2.1 \times 10^{-9}$. Also, the black points correspond to the values of the inflaton for which $N_e=49$ (upper curve) and $N_e=60$ (lower curve); the corresponding predictions for $n_s$ and $r$ are also shown.
   }\label{potential}
\end{figure}

In Fig.~\ref{potential}~\,$\mathcal{U}$ is plotted as a function of $\omega$ for $\beta=-20\bp^2$. One can note that the values of the energy density, $\sim \mathcal{U}$, are well below the Planck scale, the maximal cutoff of the theory (as Einstein's gravity breaks down at that scale). Therefore, if the UV completion of the present model occurs at energies not much smaller than $M_P$, the effective field theory approach used here is expected to provide a good approximation for the observable quantities.

\begin{figure}[t]
\begin{center}
 \includegraphics[scale=0.41]{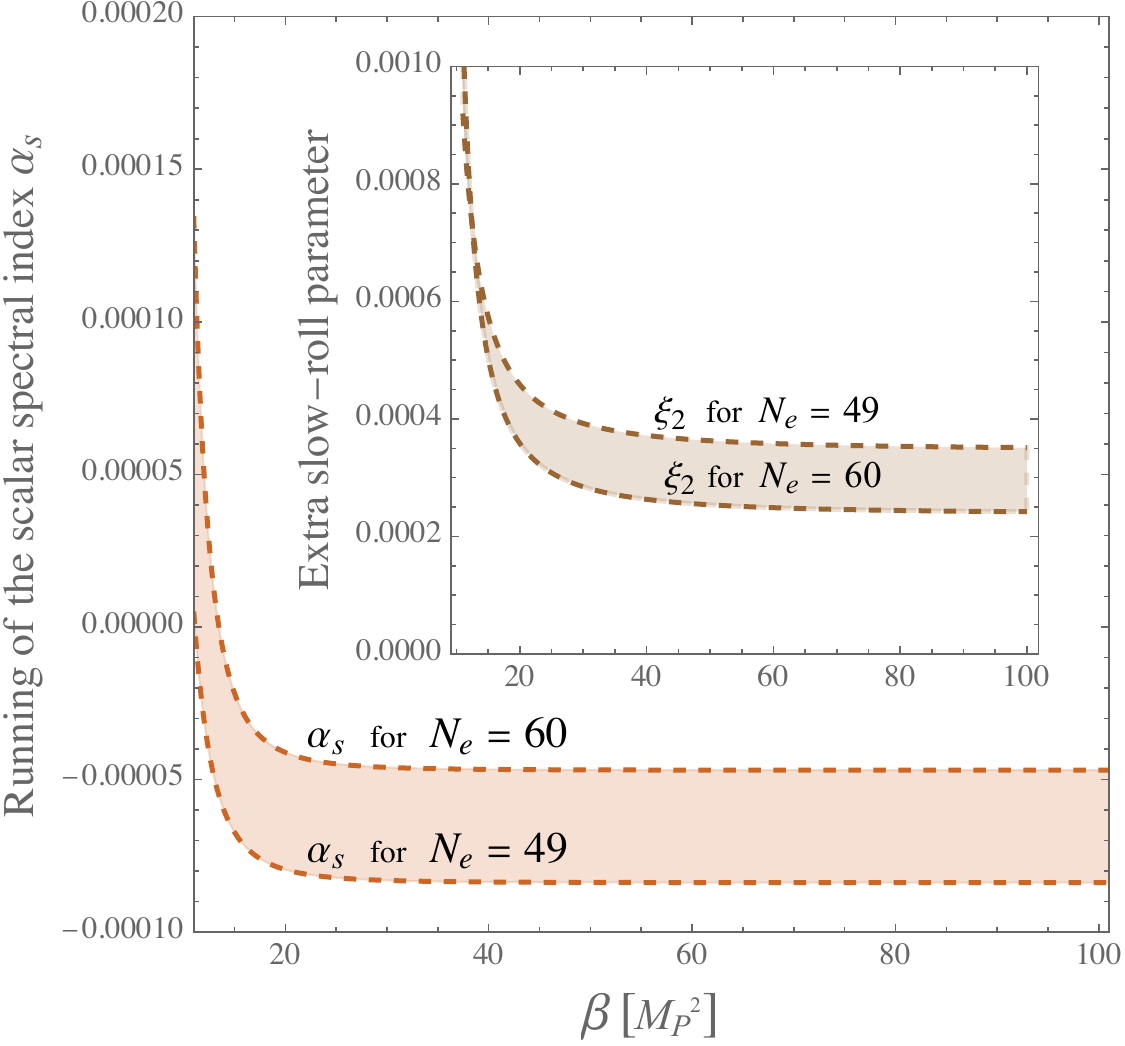} 
 \end{center} 
  \vspace{-0.5cm}
   \caption{\em Running $\alpha_s$ of the scalar spectral index as a function of $\beta$. The index shows the extra slow-roll parameter $\xi_2$. 
   }\label{running} 
\end{figure}

In Fig.~\ref{running} the predictions for $\alpha_s$ (and for $\xi_2$ in the inset) as a function of $\beta$ are shown. The predicted $\alpha_s$ is so small to be within approximately the $1\sigma$ bounds from both~\cite{ACT:2025tim} and~\cite{Planck:2018vyg}. The inset of Fig.~\ref{running} shows that the extra slow-roll parameter $\xi_2$ remains small for all relevant values of $\beta$.
 
The values of $\beta$ favored  by ACT~\cite{ACT:2025fju,ACT:2025tim} are compatible with the reheating mechanisms previously proposed in~\cite{Salvio:2022suk} up to a reheating temperature above the electroweak scale. Such efficient reheating can take place even without including extra degrees of freedom besides the Standard Model ones if one just adds to the action the following non-minimal coupling: 
 \be S_{\rm nm}=\int \sqrt{-g}\frac{\xi \phi^2}{2} {\cal R}. \label{Snm} 
  \ee 
where $\phi$ is the (canonically normalized) Higgs field in the unitary gauge and $\xi$ is a dimensionless coefficient. This is true even for values of $\xi$ that are not large, guaranteeing perturbative unitarity up to scales close to $\bp$, where an ultraviolet completion of the effective field theory we are considering should be unavoidably taken into account. Indeed, the term~(\ref{Snm}) allows for the decay of the pseudoscalaron into two Higgs bosons after inflation, leading to a reheating temperature  $T_{\rm RH}\gtrsim 10^9 |\xi|$~GeV (in the minimal setup considered here  and for $|\beta| \gtrsim \bp^2$)~\cite{Salvio:2022suk}. Note that for $|\xi|\gtrsim 1$ the reheating temperature is very large, $T_{\rm RH}\gtrsim 10^9$~GeV, and even for a very small $|\xi|$ the reheating temperature remains much larger than what is required by observations.

The term in~(\ref{Snm}) is known to be generated by quantum corrections and, therefore, it is more natural to include it. However, quantum corrections in a perturbative setup are loop suppressed and, therefore, one does not expect a large $\xi$.


Higgs inflation in both metric gravity (when ${\cal A}_{\mu~\sigma}^{~\,\rho}$ equals the Levi-Civita connection) and metric-affine gravity  typically uses $\xi\gg1$~\cite{Shaposhnikov:2020gts} to flatten and suppress the Higgs potential at inflationary scales. As a result, for values of $\xi$ that are sufficiently small (but sufficiently large to have a large-enough $T_{\rm RH}$) the dynamics of the Higgs during inflation is  negligible in the minimal setup considered here. In some extended models, however, Higgs inflation can take place even for $\xi=0$~\cite{Langvik:2020nrs}.


\section{Conclusions}

According to the new measurements and constraints provided by ACT~\cite{ACT:2025fju,ACT:2025tim} the class of observationally favored inflationary models has significantly changed compared to the previous Planck and BK18 determinations~\cite{Ade:2015lrj,BICEP:2021xfz}.

 In this work it has been shown that a simple explanation of the ACT determinations may be provided if the inflaton is a component of a dynamical connection that is independent of the metric. In this way not only gravity but also inflation has a geometrical origin. This conclusion has been reached by performing a detailed analysis of the inflationary predictions of the minimal model of this sort, whose inflationary action appears in Eq.~(\ref{SI}), including $n_s$, $r$, $P_R$ and $\alpha_s$, and by comparing them with the observational constraints. 
 
 Furthermore, here this minimal model has been shown to provide an inflationary attractor:  arbitrary initial values of the kinetic energy density are dynamically attracted down to negligible values compared to the potential energy density in homogeneous and isotropic metrics, for $\omega$ on the inflationary plateau.
 
 An interesting aspect is that the new data favor values of $\sqrt{|\beta|}$ closer to the other dimensionful quantity in the model, $\bp$. This appears to be a more natural setting. For these values of $\beta$, however, it is still easy to reheat the universe up to a temperature much larger than the electroweak scale if a non-minimal coupling of the Higgs is introduced, Eq.~(\ref{Snm}). Such term is generically induced by perturbative quantum corrections, providing a value of $\xi$ typically large enough to establish such reheating, but  small enough to neglect the dynamics of the Higgs during inflation in the minimal setup considered here. 
 
  The values of $\alpha_s$ predicted by the model are very small: as illustrated in Fig.~\ref{running}, any evidence for $\alpha_s$ significantly above the $10^{-4}$ order of magnitude would falsify the model. 
  
Interestingly, the requirement that the two scales of the model with mass dimension, $M_P$ and $|\sqrt{\beta}|$, be close to each other (which appears natural)  favours values of $n_s$ close to the upper observational bound. This can be tested by future experiments such as LiteBIRD~\cite{LiteBIRD:2022cnt}.


\begin{thebibliography}{9}

\bibitem{ACT:2025fju}
T.~Louis \textit{et al.} [ACT],
``The Atacama Cosmology Telescope: DR6 Power Spectra, Likelihoods and $\Lambda$CDM Parameters,''
[\href{http://arxiv.org/abs/2503.14452}{arXiv:2503.14452}]

\bibitem{ACT:2025tim}
E.~Calabrese \textit{et al.} [ACT],
``The Atacama Cosmology Telescope: DR6 Constraints on Extended Cosmological Models,''
[\href{http://arxiv.org/abs/2503.14454}{arXiv:2503.14454}].

\bibitem{Ade:2015lrj}
  P.~A.~R.~Ade {\it et al.} [Planck Collaboration],
  ``Planck 2015 results. XX. Constraints on inflation,''
  Astron.\ Astrophys.\  {\bf 594} (2016) A20 
  doi:10.1051/0004-6361/201525898
 [\href{http://arxiv.org/abs/1502.02114}{arXiv:1502.02114}].
  Y.~Akrami {\it et al.} [Planck Collaboration],
  ``Planck 2018 results. X. Constraints on inflation,''
  Astron. Astrophys. \textbf{641} (2020), A10
  doi:10.1051/0004-6361/201833887
 [\href{http://arxiv.org/abs/1807.06211}{arXiv:1807.06211}].
  
\bibitem{BICEP:2021xfz}
P.~A.~R.~Ade \textit{et al.} [BICEP and Keck],
``Improved Constraints on Primordial Gravitational Waves using Planck, WMAP, and BICEP/Keck Observations through the 2018 Observing Season,''
Phys. Rev. Lett. \textbf{127} (2021) no.15, 151301
doi:10.1103/PhysRevLett.127.151301
[\href{http://arxiv.org/abs/2110.00483}{arXiv:2110.00483}].

\bibitem{DESI:2024uvr}
A.~G.~Adame \textit{et al.} [DESI],
``DESI 2024 III: baryon acoustic oscillations from galaxies and quasars,''
JCAP \textbf{04} (2025), 012
doi:10.1088/1475-7516/2025/04/012
[\href{http://arxiv.org/abs/2404.03000}{arXiv:2404.03000}].
A.~G.~Adame \textit{et al.} [DESI],
``DESI 2024 VI: cosmological constraints from the measurements of baryon acoustic oscillations,''
JCAP \textbf{02} (2025), 021
doi:10.1088/1475-7516/2025/02/021
[\href{http://arxiv.org/abs/2404.03002}{arXiv:2404.03002}].

\bibitem{Kallosh:2025rni}
R.~Kallosh, A.~Linde and D.~Roest,
``A simple scenario for the last ACT,''
[\href{http://arxiv.org/abs/2503.21030}{arXiv:2503.21030}].

\bibitem{Aoki:2025wld}
S.~Aoki, H.~Otsuka and R.~Yanagita,
``Higgs-Modular Inflation,''
[\href{http://arxiv.org/abs/2504.01622}{arXiv:2504.01622}].

\bibitem{Berera:2025vsu}
A.~Berera, S.~Brahma, Z.~Qiu, R.~O.~Ramos and G.~S.~Rodrigues,
``The early universe is $\textit{ACT}$-ing $\textit{warm}$,''
[\href{http://arxiv.org/abs/2504.02655}{arXiv:2504.02655}].

\bibitem{Dioguardi:2025vci}
C.~Dioguardi, A.~J.~Iovino and A.~Racioppi,
``Fractional attractors in light of the latest ACT observations,''
[\href{http://arxiv.org/abs/2504.02809}{arXiv:2504.02809}].

\bibitem{Brahma:2025dio}
S.~Brahma and J.~Calder\'on-Figueroa,
``Is the CMB revealing signs of pre-inflationary physics?,''
[\href{http://arxiv.org/abs/2504.02746}{arXiv:2504.02746}].

\bibitem{Gialamas:2025kef}
I.~D.~Gialamas, A.~Karam, A.~Racioppi and M.~Raidal,
``Has ACT measured radiative corrections to the tree-level Higgs-like inflation?,''
[\href{http://arxiv.org/abs/2504.06002}{arXiv:2504.06002}].

\bibitem{Baldazzi:2021kaf}
A.~Baldazzi, O.~Melichev and R.~Percacci,
``Metric-Affine Gravity as an effective field theory,''
Annals Phys. \textbf{438} (2022), 168757
doi:10.1016/j.aop.2022.168757
[\href{http://arxiv.org/abs/2112.10193}{arXiv:2112.10193}].

\bibitem{Pradisi:2022nmh}
G.~Pradisi and A.~Salvio,
``(In)equivalence of metric-affine and metric effective field theories,''
Eur. Phys. J. C \textbf{82} (2022) no.9, 840
doi:10.1140/epjc/s10052-022-10825-9
[\href{http://arxiv.org/abs/2206.15041}{arXiv:2206.15041}].

\bibitem{Salvio:2022suk}
A.~Salvio,
``Inflating and reheating the Universe with an independent affine connection,''
Phys. Rev. D \textbf{106} (2022) no.10, 103510
doi:10.1103/PhysRevD.106.103510
[\href{http://arxiv.org/abs/2207.08830}{arXiv:2207.08830}].

\bibitem{DiMarco:2023ncs}
A.~Di Marco, E.~Orazi and G.~Pradisi,
``Einstein\textendash{}Cartan pseudoscalaron inflation,''
Eur. Phys. J. C \textbf{84} (2024) no.2, 146
doi:10.1140/epjc/s10052-024-12482-6
[\href{http://arxiv.org/abs/2309.11345}{arXiv:2309.11345}].


\bibitem{Karananas:2025xcv}
G.~K.~Karananas,
``Geometrical origin of inflation in Weyl-invariant Einstein-Cartan gravity,''
Phys. Lett. B \textbf{862} (2025), 139343
doi:10.1016/j.physletb.2025.139343
[\href{http://arxiv.org/abs/2501.16416}{arXiv:2501.16416}].

\bibitem{indices}Greek indices are lowered and raised by the metric $g_{\mu\nu}$, for which the mostly plus convention is assumed here. 

\bibitem{Hojman:1980kv}
R.~Hojman, C.~Mukku and W.~A.~Sayed,
``Parity violation in metric torsion theories of gravitation,''
Phys. Rev. D \textbf{22} (1980), 1915-1921
doi:10.1103/PhysRevD.22.1915.

\bibitem{Nelson:1980ph}
P.~C.~Nelson,
``Gravity With Propagating Pseudoscalar Torsion,''
Phys. Lett. A \textbf{79} (1980), 285
doi:10.1016/0375-9601(80)90348-5.

\bibitem{Holst:1995pc}
S.~Holst,
``Barbero's Hamiltonian derived from a generalized Hilbert-Palatini action,''
Phys. Rev. D \textbf{53} (1996), 5966-5969
doi:10.1103/PhysRevD.53.5966
[\href{http://arxiv.org/abs/gr-qc/9511026}{arXiv:gr-qc/9511026}].

\bibitem{Racioppi:2024zva}
A.~Racioppi and A.~Salvio,
``Natural metric-affine inflation,''
JCAP \textbf{06} (2024), 033
doi:10.1088/1475-7516/2024/06/033
[\href{http://arxiv.org/abs/2403.18004}{arXiv:2403.18004}].

\bibitem{Immirzi:1996di}
G.~Immirzi,
``Real and complex connections for canonical gravity,''
Class. Quant. Grav. \textbf{14} (1997), L177-L181
doi:10.1088/0264-9381/14/10/002
[\href{http://arxiv.org/abs/gr-qc/9612030}{arXiv:gr-qc/9612030}].

\bibitem{Immirzi:1996dr}
G.~Immirzi,
``Quantum gravity and Regge calculus,''
Nucl. Phys. B Proc. Suppl. \textbf{57} (1997), 65-72
doi:10.1016/S0920-5632(97)00354-X
[\href{http://arxiv.org/abs/gr-qc/9701052}{arXiv:gr-qc/9701052}].

\bibitem{Hecht:1996np}
R.~D.~Hecht, J.~M.~Nester and V.~V.~Zhytnikov,
``Some Poincare gauge theory Lagrangians with well posed initial value problems,''
Phys. Lett. A \textbf{222} (1996), 37-42
doi:10.1016/0375-9601(96)00622-6.

\bibitem{BeltranJimenez:2019hrm}
J.~Beltr\'an Jim\'enez and F.~J.~Maldonado Torralba,
``Revisiting the stability of quadratic Poincar\'e gauge gravity,''
Eur. Phys. J. C \textbf{80} (2020) no.7, 611
doi:10.1140/epjc/s10052-020-8163-8
[\href{http://arxiv.org/abs/1910.07506}{arXiv:1910.07506}].

\bibitem{Gialamas:2022xtt}
I.~D.~Gialamas and K.~Tamvakis,
``Inflation in metric-affine quadratic gravity,''
JCAP \textbf{03} (2023), 042
doi:10.1088/1475-7516/2023/03/042
[\href{http://arxiv.org/abs/2212.09896}{arXiv:2212.09896}].

\bibitem{HomIso}
Homogeneity and  isotropy is implied by three-dimensional translational and rotational symmetries.

\bibitem{Linde:1985ub}
  A.~D.~Linde,
  ``Initial Conditions For Inflation,''
  Phys.\ Lett.\ B {\bf 162} (1985) 281
  doi:10.1016/0370-2693(85)90923-2.
   %
\bibitem{Salvio:2015kka}
A.~Salvio and A.~Mazumdar,
``Classical and Quantum Initial Conditions for Higgs Inflation,''
Phys. Lett. B \textbf{750} (2015), 194-200
doi:10.1016/j.physletb.2015.09.020
[\href{http://arxiv.org/abs/1506.07520}{arXiv:1506.07520}].

\bibitem{Salvio:2017oyf}
A.~Salvio,
``Initial Conditions for Critical Higgs Inflation,''
Phys. Lett. B \textbf{780} (2018), 111-117
doi:10.1016/j.physletb.2018.03.009
[\href{http://arxiv.org/abs/1712.04477}{arXiv:1712.04477}].
  
\bibitem{Starobinsky:1980te}
A.~A.~Starobinsky,
``A New Type of Isotropic Cosmological Models Without Singularity,''
Phys. Lett. B \textbf{91} (1980), 99-102
doi:10.1016/0370-2693(80)90670-X. 


\bibitem{Planck:2018vyg}
N.~Aghanim \textit{et al.} [Planck],
``Planck 2018 results. VI. Cosmological parameters,''
Astron. Astrophys. \textbf{641} (2020), A6
[erratum: Astron. Astrophys. \textbf{652} (2021), C4]
doi:10.1051/0004-6361/201833910
[\href{http://arxiv.org/abs/1807.06209}{arXiv:1807.06209}].

\bibitem{Shaposhnikov:2020gts}
M.~Shaposhnikov, A.~Shkerin, I.~Timiryasov and S.~Zell,
``Higgs inflation in Einstein-Cartan gravity,''
JCAP \textbf{02} (2021), 008
[erratum: JCAP \textbf{10} (2021), E01]
doi:10.1088/1475-7516/2021/10/E01
[\href{http://arxiv.org/abs/2007.14978}{arXiv:2007.14978}].

\bibitem{Langvik:2020nrs}
M.~L\r{a}ngvik, J.~M.~Ojanper\"a, S.~Raatikainen and S.~Rasanen,
``Higgs inflation with the Holst and the Nieh\textendash{}Yan term,''
Phys. Rev. D \textbf{103} (2021) no.8, 083514
doi:10.1103/PhysRevD.103.083514
[\href{http://arxiv.org/abs/2007.12595}{arXiv:2007.12595}].

\bibitem{LiteBIRD:2022cnt}
E.~Allys \textit{et al.} [LiteBIRD],
``Probing Cosmic Inflation with the LiteBIRD Cosmic Microwave Background Polarization Survey,''
[\href{https://arxiv.org/abs/2202.02773}{arXiv:2202.02773}].

\end{thebibliography}
\end{document}